\documentclass[prletter,showpacs,twocolumn,typeset,superscriptaddress]{revtex4}

\usepackage{amsfonts}
\usepackage{amsmath}
\usepackage{amssymb}
\usepackage{graphicx}
\usepackage{epsfig}
\usepackage{citesort}

\usepackage{dcolumn}
\usepackage{bm}

\begin{document}

\preprint{APS}

\title{Quantum phase transitions in an effective Hamiltonian: fast and slow
systems}
\author{Isabel Sainz}
\affiliation{School of Information and Communication Technology,
Royal Institute of Technology (KTH), Electrum 229, SE-164 40 Kista,
Sweden}
\author{A. B. Klimov}
\affiliation{Departamento de F\'{\i}sica, Universidad de
Guadalajara, Revoluci\'{o}n 1500, 44420 Guadalajara, Jalisco,
Mexico.}
\author{Luis Roa}
\affiliation{Center for Quantum Optics and Quantum Information,
Departamento de F\'{\i}sica, Universidad de Concepci\'{o}n, Casilla
160-C, Concepci\'{o}n, Chile.}

\date{\today}

\begin{abstract}
An effective Hamiltonian describing interaction between generic 
\textit{fast} and a \textit{slow} systems is obtained in 
the strong interaction limit. The result is applied for studying
the effect of quantum phase transition as a bifurcation of the ground
state of the \textit{slow subsystem} in the thermodynamic limit.
Examples as atom-field and atom-atom interactions are analyzed in detail. 
\end{abstract}

\pacs{42.50.Ct, 42.50.Hz, 42.50.Fx}

\maketitle

\section{Introduction}

Frequently, in the process of interaction between two quantum
systems, only one of them can be detected experimentally. In this
case, a variety of physical effects appear in the process of such
interaction which can be described in terms of an effective
Hamiltonian corresponding to the observed system. The simplest
example of such a situation arises when a \textit{fast system}
interacts with a \textit{slow system}. Then, the \textit{fast
system} can be adiabatically eliminated and the \textit{slow system}
is described by an effective Hamiltonian. These considerations were
assumed in the famous Born-Oppenheimer approximation. A regular
approach to the quantum dynamics of the observed system is provided
by the Lie transformation method \cite{Lie,SR}. The advantage of
this method consists in the possibility of varying the system's
parameters, changing relations between them, which allows us to
describe different physical regimes using the same mathematical
tool. In particular, such an important example as expansion on the
resonances in quantum systems not preserving the number of
excitations can be obtained \cite{RWA}. In this case a generic
Hamiltonian governing interaction of two subsystems beyond the
Rotating Wave Approximation (RWA) can be represented as a series in
operators describing all possible transitions in the system.

Several interesting features appearing in the process of interaction
of quantum systems can be realized by studying evolution of only two
generic quantum system with one quantum channel. Even in such a
simple case we may discriminate at least three interesting limits:
a) when the interaction constant $g$ is much higher than the
characteristic frequencies of both interacting systems; b) when $g$
is smaller than the frequencies of the systems and c) when $g$ is
higher than the frequency of one system but smaller that the
frequency of the other one.

The a) case of very strong coupling should be studied carefully,
because using the expansion parameter like an interaction constant
over a characteristic frequency could be quite tricky. For instance,
the type of the spectrum corresponding to the non-perturbed and to
the perturbed systems can be different: either continuous or
discrete.

The b) case corresponds to a situation where the resonance expansion
is applicable. This particular case leads to dispersive-like
interactions \cite{LRAK}. As it was shown in Ref. \cite{RWA}, the
evolution is governed by an effective Hamiltonian describing a
certain resonant interaction and the representation space of the
total system can be always divided into (almost) invariant
subspaces.

The last case c) possesses a peculiar property: besides of finding a
corresponding effective Hamiltonian, we can also project it out to
the lower energy state of the fast system, which would never get
excited under given relations between the system's parameters, and
thus, describe an effective dynamics of the slow system in the limit
of strong interaction. It is well known that in this regime such an
interesting effect as Quantum Phase Transitions may occur.

The quantum phase transitions (QPT) are a common feature of
non-linear quantum systems. Such transitions occur at zero
temperature and are associated with an abrupt change in the ground
state structure. QPT are related to singularities in the energy
spectrum and, at the critical points defining QPT, the ground state
energy is a non-analytic function of the system's parameters
\cite{iachello}. Qualitatively, for a wide class of quantum systems,
several important properties of QPT can be studied in the so-called
thermodynamic (semiclassical) limit \cite{TL,LM}. Then, QPT can be
analyzed in terms of a classical effective potential energy surface
\cite{dicke}. In this language QPT are related to the appearance of
a new classical separatrix when the coupling parameters acquire
certain values. According to the standard semiclassical quantization
scheme and the correspondence principle, the energy density is
proportional to the classical period of motion, diverging on the
separatrix, which explains a high density of quantum states at the
critical points.

In this article we study effective Hamiltonians describing evolution
of a generic quantum system $X$ interacting with a quantum system
$Y$ in the case where the characteristic frequency of the system $X$
is essentially lower than the corresponding frequency of the system
$Y$, $\omega_{X}\ll\omega_{Y}$, and the interaction constant $g$
satisfies the strong coupling condition: $\omega _{X}\lesssim
g\ll\omega_{Y}$. We show that, depending on the type of interaction
and the nature of quantum systems different physical situations take
place, but generically such effective Hamiltonians describe Quantum
Phase Transitions in the \textit{slow system}.

\section{Effective Hamiltonian}

Let us consider the following generic Hamiltonian describing an
interaction between two quantum systems:
\begin{equation}
H=\omega _{1}X_{0}+\omega _{2}Y_{0}+g\left( X_{+}+X_{-}\right) \left(
Y_{+}+Y_{-}\right) ,  \label{Hin}
\end{equation}%
where $X_{0}$ and $Y_{0}$ are the free Hamiltonians of the $X$ and
$Y$ systems respectively, and such that $ \omega_{1}\ll \omega_{2}$.
The above Hamiltonian does not preserve the \textit{total excitation
number operator} $N=X_{0}+Y_{0}$ and, in the limit
$\omega_{1},\omega_{2}\gg g$, leads to the appearance of
\textit{multiphoton-type} interactions of the form
$X_{+}^{n}Y_{-}^{m}$ which, under certain physical conditions on the
frequencies $\omega _{1,2}$, describe \textit{resonant} transitions
between energy levels of the whole system (see \cite{RWA} and
references therein).

The raising-lowering operators $X_{\pm}$, $Y_{\pm}$ describe
transitions between energy levels of the systems $X$ and $Y$
respectively and consequently obey the following commutation
relations:
\begin{equation}
\left[ X_{0},X_{\pm }\right] =\pm X_{\pm },\,\quad \left[ Y_{0},Y_{\pm }%
\right] =\pm Y_{\pm }.  \label{CR}
\end{equation}
We do not impose any condition on the commutators between transition
operators, which are generally some functions of diagonal operators
and of some integrals of motion $\left[ N_{1},X_{0}\right] =\left[
N_{2},Y_{0}\right]=0$:
\begin{subequations}
\begin{eqnarray}
\left[X_{+},X_{-}\right]&=&\nabla_{X_{0}}\phi_{1}(X_{0},N_{1}),\label{PZ1}\\
\left[Y_{+},Y_{-}\right]&=&\nabla_{Y_{0}}\phi
_{2}(Y_{0},N_{2}),\label{PZ2}
\end{eqnarray}
\label{PZ}
\end{subequations}%
where $\phi_{1}(X_{0},N_{1})=X_{+}X_{-}$ and $\phi
_{2}(Y_{0},N_{2})=Y_{+}Y_{-} $ are some polynomials of $X_{0}$ and
$Y_{0}$ respectively (from now on we omit the dependence on
integrals $N_{1,2}$ in the arguments) and $\nabla_{z}\phi (z)=\phi
(z)-\phi (z+1)$. The objects ($X_{0},X_{\pm }$) and ($Y_{0},Y_{\pm
}$) are known as polynomial deformed algebras $sl_{pd}(2,R)$
\cite{deform}.

Now, we will be interested in the limit where the \textit{slow
system} frequency is less than/or of the order of the coupling
constant, $\omega _{1}\lesssim g\ll \omega _{2}$. Following the
method described in Ref. \cite{SR} we can adiabatically remove all
the terms that contain the fast system's transition operators,
$Y_{\pm}$. In particular, the counter-rotating term
$X_{+}Y_{+}+X_{-}Y_{-}$ and the rotating term
$X_{+}Y_{-}+X_{-}Y_{+}$ can be eliminated from the Hamiltonian
(\ref{Hin}) by a subsequent application of the following Lie-type
transformations:
\begin{subequations}
\begin{eqnarray}
U_{1} &=&\exp \left[ \varepsilon \left( X_{+}Y_{+}-X_{-}Y_{-}\right) \right]
,  \label{U1} \\
U_{2} &=&\exp \left[ \epsilon \left( X_{+}Y_{-}-X_{+}Y_{-}\right) \right] ,
\label{U2}
\end{eqnarray}
\label{U}
\end{subequations}%
where the small parameters, $\varepsilon$ and $\epsilon$, are
defined by
\begin{equation}
\varepsilon =\frac{g}{\omega _{2}+\omega _{1}}\ll
1\qquad\text{and}\qquad\epsilon =\frac{g }{\omega _{2}-\omega
_{1}}\ll 1.  \label{epcilon}
\end{equation}

The transformations (\ref{U1}) and (\ref{U2}) generate different
kinds of terms: such as $X_{\pm }^{n}Y_{\pm }^{k}+h.c.$, $X_{\pm
}^{n}Y_{\mp }^{k}+h.c.$, $Y_{\pm }^{n}+h.c.$, and $X_{\pm}^{n}+h.c.$
with coefficients depending on $X_{0}$ and $Y_{0}$. Under the
condition $\omega_{1},g\ll \omega _{2}$ all the rapidly oscillating
terms, i.e. those containing powers of $Y_{\pm },$ can be removed by
applying transformations similar to (\ref{U}), with properly chosen
parameters. Then, the effective Hamiltonian is diagonal for the
operators of the $Y$ system. The result can be expressed as a power
series of the single parameter $\delta=g/\omega_{2}\ll 1$.

It is worth noting that it is not enough that $\delta $ be a small
parameter for the formal expansion in (\ref{U}) (and the subsequent
transformations). A balance is necessary between the
\textit{effective dimensions} of the subsystems and $\delta $. The
\textit{effective dimensions} of the system depend on the order of
the polynomials $\phi_{1,2}$, and on the powers of the elements
$X_{\pm,0}$ and $Y_{\pm,0}$ involved in each transformation. It was
shown before \cite{RWA}, that the powers of the small parameters are
increasing faster than the powers of $X_{\pm,0}$ and $Y_{\pm,0}$,
which implies that we can focus on the \textit{effective dimensions}
introduced with (\ref{U}).

Taking into account the above mentioned considerations, keeping only
terms up to third order in $\delta$ and disregarding small
corrections to the effective transition frequencies, we arrive at
the following effective Hamiltonian:
\begin{widetext}
\begin{eqnarray}
H_{eff}&=&\omega _{1}X_{0}+\omega _{2}Y_{0}-2\omega _{1}\delta
^{2}\nabla _{x,-y}\Phi (X_{0},Y_{0}+1)+g\delta \nabla _{y}\phi
_{y}(Y_{0})\left(
X_{+}+X_{-}\right) ^{2}  \nonumber \\
&&+\frac{1}{2}g\delta ^{3}\nabla _{y}\left( \phi _{y}(Y_{0})\nabla
_{y}^{2}\phi _{y}(Y_{0}-1)\right) \left( X_{+}+X_{-}\right) ^{4},
\label{Heff1}
\end{eqnarray}
\end{widetext}
where
\begin{equation}
\Phi (X_{0},Y_{0}+1)=\nabla _{X_{0},Y_{0}}\left[ \phi _{1}\left(
X_{0}\right) \phi _{2}(Y_{0})\right] ,  \label{phi}
\end{equation}
and the generalized displacement operators are defined as
\begin{equation}
\nabla_{mX_{0},nY_{0}}f(X_{0},Y_{0})=f(X_{0},Y_{0})-f(
X_{0}+m,Y_{0}+n), \nonumber
\end{equation}
for $m$ and $n$ integers.

Because the effective Hamiltonian (\ref{Heff1}) is diagonal for the
operators of the $Y$ \textit{fast system}, we may project it out
onto a minimal energy eigenstate of the $Y$ system, $|\psi
_{0}\rangle _{Y}$, substituting $Y_{0}$ by its eigenvalue $y_{0}$:
$Y_{0}|\psi_{0}\rangle_{Y}=y_{0}|\psi_{0}\rangle_{Y}.$

The first order effect then comes from the term
$\sim\left(X_{+}+X_{-}\right)^{2}$, while the term
$\sim\left(X_{+}+X_{-}\right)^{4}$ defines a \textit{fine structure}
of the \textit{effective potential}, obtained after projecting the
effective Hamiltonian (\ref{Heff1}) onto the state
$|\psi_{0}\rangle_{Y}$.

It is important to stress that, although $\delta$ is a small
parameter, the effect of the terms $\sim\delta^{n}$, $n\geq 1$,
could be in principle comparable with the main diagonal term
$\omega_{1}X_{0}$, especially if the algebra of $X$ operators
describe a \textit{big subsystem}, i.e., large spin or big photon
number. In this case non-trivial effects such as QPT may occur.

Now, we may proceed with analysis of the effective Hamiltonian (\ref{Heff1})
in the thermodynamic limit, focusing on the possible bifurcation of the
ground state.

\section{Examples}

\subsection{Atom-field interaction (Dicke model)}

The Hamiltonian governing the evolution of $A$ symmetrically prepared
two-level atoms interacting with a single mode of quantized field has the
form
\begin{equation}
H=\omega _{1}\hat{n}+\omega _{2}S_{z}+g\left( S_{+}+S_{-}\right)
\left( a^{\dagger }+a\right),  \label{HaS}
\end{equation}
where $\hat{n}=a^{\dagger}a$ and $S_{z,\pm }$ are generators of the
$(A+1)$-dimensional representation of the $su(2)$ algebra.

\subsubsection{Effective field dynamics}

First let us suppose that the atoms form a \textit{fast subsystem}
so that,
\begin{equation}
X_{0}=\hat{n},\hspace{0.1in}X_{+}=a^{\dagger
},\hspace{0.1in}X_{-}=a,\hspace{0.1in}\;Y_{0}=S_{z},\hspace{0.1in}Y_{\pm}=S_{\pm},
\nonumber
\end{equation}
and thus, $\phi _{y}(Y_{0})=C_{2}-S_{z}^{2}+S_{z}$ and $\phi
_{x}(X_{0})=\hat{n}$, where $C_{2}=A/2(A/2+1)$ is the eigenvalue of
the Casimir operator of the $su(2)$ algebra (integral of motion
corresponding to the atomic subsystem).

Projecting the effective Hamiltonian onto the minimum energy state
of the atomic system $|0\rangle_{at}$, so that $y_{0}=-A/2$, we
obtain the following effective Hamiltonian for the field mode:
\begin{equation}
H_{eff}=\tilde{\omega}_{1}\hat{n}-Ag\delta\left(a+a^{\dagger}\right)^{2}+gA\delta^{3}\left(a+a^{\dagger}\right)^{4},
\label{Heff_f}
\end{equation}
where $\tilde{\omega}_{1}=\omega_{1}(1-2A\delta^{2}) $.

Rewriting (\ref{Heff_f}) in terms of position and momentum
operators,
\begin{equation}
H_{eff}=\frac{\tilde{\omega}_{1}}{2}(p^{2}+x^{2})-2Ag\delta
x^{2}+4gA\delta^{3}x^{4}, \nonumber
\end{equation}
we immediately detect that QPT in this case is related to the
bifurcation of the effective potential
$U(x)=(\tilde{\omega}_{1}/2-2Ag\delta)x^{2}+4gA\delta^{3}x^{4}$ from
a single minimum at $A<\tilde{\omega}_{1}/(4g\delta )$ to a double
well structure at $A>\tilde{\omega}_{1}/(4g\delta )$. The physical
effect associated with this QPT consists of a spontaneous generation
of photons in the field mode. In some sense, the virtual photons,
always presented in the Dicke model (\ref{HaS}), are
\textit{condensed} into the real photons after crossing the critical
point $A=\tilde{\omega}_{1}/(4g\delta )$. It is worth noting that
this does not happen if the RWA is applied to (\ref{HaS}).

\subsubsection{Effective atomic dynamics}

In the opposite case, when the atoms form a \textit{slow subsystem}
we have
\begin{equation}
X_{0}=S_{z},\hspace{0.1in}X_{\pm}=S_{\pm},\hspace{0.1in}Y_{0}=\hat
n,\hspace{0.1in}Y_{+}=a^{\dagger },\hspace{0.1in}Y_{-}=a.\nonumber
\end{equation}
Projecting the effective Hamiltonian onto the minimum energy state
of the field mode $|0\rangle_{f}$, so that $y_{0}=0$, the effective
Hamiltonian acquires the form
\begin{equation}
H_{eff}=\tilde{\omega}_{1}S_{z}-4g\delta S_{x}^{2}+2\omega _{2}\delta
^{2}S_{z}^{2}+16g\delta ^{3}\{S_{x}^{2},S_{z}\},  \label{Heff_at}
\end{equation}
where
$\tilde{\omega}_{1}=\omega_{1}-2\omega_{2}\delta^{2}-16g\delta^{3}$.

For our analysis it is convenient to perform a $\pi/2$ rotation in
(\ref{Heff_at}) around axis $y$, transforming the (\ref{Heff_at})
Hamiltonian into
\begin{equation}
\tilde{H}_{eff}=-\tilde{\omega}_{1}S_{x}-4g\delta S_{z}^{2}+2\omega
_{2}\delta ^{2}S_{x}^{2}-16g\delta ^{3}\{S_{z}^{2},S_{x}\}.  \label{Heff_at1}
\end{equation}

In the thermodynamical limit we may replace the atomic operators by
the corresponding classical vectors over the two-dimensional sphere,
i.e.,
\begin{equation}
S_{z}\rightarrow\frac{A}{2}\cos\theta,\hspace{0.1in}S_{x}\rightarrow
\frac{A}{2}\sin\theta\cos\phi,\hspace{0.1in}S_{y}\rightarrow\frac{A}{2}\sin\phi\sin\theta,\nonumber
\end{equation}
and thus rewrite the effective Hamiltonian (\ref{Heff_at1}) as a
classical Hamiltonian function,
\begin{eqnarray}
H_{cl}&=&-\frac{A}{2}(\tilde{\omega}_{1}\cos\phi\sin\theta+2Ag\delta\cos^{2}\theta\nonumber\\
&&-A\omega_{2}\delta^{2}\cos^{2}\phi\sin^{2}\theta\nonumber\\
&&+4A^{2}g\delta^{3}\cos\theta\cos\phi\sin^{2}\theta).
\label{Hcl_at}
\end{eqnarray}
The first two terms in the above expression describe the
thermodynamical limit of the Lipkin-Meshkov model \cite{LM} and
determine the critical point of QPT,
$\xi=4Ag\delta/\tilde{\omega}_{1}=1$, which again is related to the
bifurcation of the ground state: a single minimum at $\sin\phi=0$,
$\cos\theta_{\ast}=0$ splits into two minima at $\sin\phi=0$,
$\cos\theta_{\ast\ast}=\pm\sqrt{1-\xi ^{-2}}$ for $\xi>1$. It is
worth noting that the global minimum of $H_{cl}$ at $\xi<1$ converts
into a local maximum for $\xi>1$, so that
$H_{cl}(\theta_{\ast\ast})<H_{cl}(\theta_{\ast})$. This means that
the atoms, initially prepared at the minimum of the Hamiltonian
function, spontaneously change their ground state energy at some
value of the system's parameters. Classically, this implies
appearance of a separatrix, which leads to the discontinuity on the
energy density spectrum in the thermodynamic limit. It is also worth
noting that there is a loss of the rotational symmetry in this
process: the new ground state is obviously not invariant under
rotations around axis $x$, while the initial ground state is clearly
invariant under $x$-rotations.


It is easy to see that the last two terms in (\ref{Hcl_at}) are of
lower order in the parameter $A\delta $ and can be neglected in the
first approximation for description of QPT at $\xi =1$.

\subsection{Spin-spin interaction}

As a second example let us consider a dipole-dipole like
interaction, that is,
\begin{equation}
X_{0}=S_{z1},\;X_{\pm }=S_{\pm 1},\quad
Y_{0}=S_{z2},\;Y_{\pm}=S_{\pm 2}.\nonumber
\end{equation}
The effective Hamiltonian for the \textit{slow spin system} (after
projecting onto the lowest state of the \textit{fast spin system}
with eigenvalue $-A_{2}/2$) takes the form similar to
(\ref{Heff_at}), with
\begin{eqnarray}
H_{eff}&=&\tilde{\omega}_{1}S_{z1}-2A_{2}g\delta
S_{x1}^{2}+2A_{2}\omega _{1}\delta ^{2}S_{z1}^{2}\nonumber\\
&&+16gA_{2}\delta ^{3}S_{x1}^{4}+24g\delta
^{3}A_{2}^{2}\{S_{x}^{2},S_{z}\}, \label{Heff_aa}
\end{eqnarray}
where $\tilde{\omega}_{1}=\omega _{1}-2A_1\omega _{1}\delta
^{2}-20g\delta^{3}A_1^{2}$. The first two terms are dominant for
$\delta \ll 1$ and describe the Lipkin-Meshkov model, so that the
critical point is reached at $\xi=4A_{2}A_{1}^{2}g\delta
/\tilde{\omega}_{1}=1$ in the thermodynamical limit. The effect of
the rest of the terms in (\ref{Heff_aa}) is negligible in the
vicinity of $\xi=1$.

\section{Conclusions}

We deduce the effective Hamiltonian of a generic slow quantum system
interacting with another fast oscillating system when the total
excitation number is not preserved. Analyzing those effective
Hamiltonians in the thermodynamic limit we have observed a
bifurcation of the ground state leading to the effect of the Quantum
Phase Transitions.

It is interesting to note that, for multidimensional systems, when
algebraically the $X$ system is a direct sum of several
non-interacting subsystems, an interesting effect of generation of
entangled states (in the non-preserving excitation case) can be
observed. Really, let us suppose that in (\ref{Hin}) $X_{0,\pm
}=X_{0,\pm 1}+X_{0,\pm 2}$, $\left[ X_{j,1},X_{j,2}\right]
=0,j=0,\pm$; then the corresponding effective Hamiltonian (up to a
first non-trivial order in $\delta $) takes the form
\begin{eqnarray}
H_{eff}&\approx&\omega_{1}(X_{0,1}+X_{0,2})+\omega_{2}Y_{0}+\nonumber\\
&&g\delta[(X_{+,1}+X_{-,1})^{2}+(X_{+,2}+X_{-,2})^{2}\nonumber\\
&&+2(X_{+,1}+X_{-,1})(X_{+,2}+X_{-,2})]\nabla_{y}\phi_{y}(Y_{0}),\nonumber
\end{eqnarray}
where we can clearly see that the last term contains the operator
product $\sim X_{+,1}X_{+,2}$ which, together with quadratic terms
in $X_{\pm,1(2)}$, implies a spontaneous generation of entangled
states of $X_{1}$ and $X_{2}$ starting from the minimum energy
state. This can be corroborated by the entangling power measure by
considering a uniform distribution of the initial factorized states
\cite{Zan}. Thus we can say that, in the regime studied here,
entanglement can be generated in a bipartite system the vicinity of
a Phase Transition.

\begin{acknowledgments}
One of the authors (I. S.) thanks STINT (Swedish Foundation for
International Cooperation in Research and Higher Education) for
support. This work was supported by Grants: CONACyT
N$^{\text{\underline{o}}}$45704, Milenio ICM P06-067-F and FONDECyT
N$^{\text{\underline{o}}}$ 1080535.
\end{acknowledgments}

\end{document}